\documentstyle[sprocl,epsfig]{article}

\newcommand{\sBe}{$^7$Be~}
\newcommand{\oB}{$^8$B~}

\def\lapprox{\mathrel{\mathop
  {\hbox{\lower0.5ex\hbox{$\sim$}\kern-0.8em\lower-0.7ex\hbox{$<$}}}}}
\def\gapprox{\mathrel{\mathop
  {\hbox{\lower0.5ex\hbox{$\sim$}\kern-0.8em\lower-0.7ex\hbox{$>$}}}}}
\def\mathrm{\mbox}
\def\eg{{\em e.g.}\/}
\def\etal{{\em et al.}\/}

\def\ie{{\em i.e.}\/}
\def\text{\textstyle}

\def\fipppep{\Phi_{\mbox{\footnotesize{pp+pep}}}}
\def\fibe{\Phi_{\mbox{\footnotesize{Be}}}}
\def\ficno{\Phi_{\mbox{\footnotesize{CNO}}}}
\def\fibecno{\Phi_{\mbox{\footnotesize{Be+CNO}}}}
\def\fib{\Phi_{\mbox{\footnotesize{B}}}}

\def\permille{$^\circ/_{\circ \circ} $}

\def\ug{$U$}
\def\N_F{$n_F$}
\def\nnu{\nu}
\def\oomega{\delta \nu}

\def\wDelta{\delta}

\def\be{\begin{equation}}
\def\ee{\end{equation}}
\def\bea{\begin{eqnarray}}
\def\eea{\end{eqnarray}}
\begin{document}
\setcounter{table}{0}

\title{Solar neutrinos: where we are and what is next?}

\author{G. FIORENTINI and B. RICCI}

\address{Dip. di Fisica, Universit\'a di Ferrara and INFN-Ferrara,
Via Paradiso 12,\\ I-44100 Ferrara 
\\E-mail: fiorentini@axpfe1.fe.infn.it, ricci@axpfe1.fe.infn.it}


\maketitle\abstracts{
We summarize the results of solar neutrino experiments and update
a solar model independent analysis of solar neutrino data.
We discuss the implications of helioseismology on solar models
and predicted solar neutrino fluxes. Finally , we discuss the potential
of new experiments for detecting specific signatures of the proposed
solutions to the solar neutrino puzzle.
}

\section{Introduction}
\label{intro}

The aim of this paper is to present the status of art
concerning  the solar neutrino physics, by  addressing the following questions:

\noindent
1) {\em What has been measured?} 
In Sec. \ref{secresult} we summarize the
 results of the five solar neutrino experiments, all  reporting
a deficit in the signal with respect the prediction of 
Standard Solar Models (SSM).

\noindent
2) {\em What  have we learnt, independently of SSMs?}
We update (Sec. \ref{secmodel}) a solar model independent analysis
of solar neutrino data
and  we show that experimental data are more and more against 
the hypothesis of standard neutrinos (\ie ~without mass, mixing, magnetic 
moments...).

\noindent
3){\em What has been calculated?}
Accurate predictions of solar neutrino fluxes
are extremely important and thus refined solar models are necessary.
These models have now to
account for several solar properties determined by means of helioseismology.
 In Sec. \ref{sechelios} we quantitatively estimate 
the accuracy of solar properties as inferred from the measured frequencies
 through the so called inversion method and discuss SSMs 
in comparison with helioseismic results.

\noindent
4){\em What is missing?}
Actually one  now needs a direct footprint of some neutrino
property, not predicted within the minimal standard model
of electroweak interaction. In this respect,
 we discuss (Sec. \ref{secfuture}) the
potential of ongoing and
 future solar neutrino experiments.

\section{Solar neutrino experiments}
\label{secresult}
So far we have results from five solar neutrino experiments, see
Table 1 for a summary and Refs.~\cite{Report,Bahcall1989,Koshiba}
for detailed reviews.

The KAMIOKANDE  (termined in 1995) and SUPERKAMIOKANDE (data taking since
April 1996) experiments 
\cite{Kamioka,Superkam}, located in the Japanese Alps,
detect  the Cerenkov light emitted by electrons that are scattered in 
the forward direction by solar neutrinos, through the reaction
\begin{equation}
	\nu+e \rightarrow \nu+e \, .
\end{equation}
 These experiments, being sensitive to 
the neutrino direction, are the prototype of neutrino telescopes and 
are the only real-time experiments so far. 
The experiments are  only sensitive to the high energy neutrinos 
from $^8$B decay.
Within the observational uncertainties,
the solar neutrino spectrum deduced from SUPERKAMIOKANDE  
(and first by KAMIOKANDE) agrees  with that of neutrinos 
from $^8$B decay in the laboratory.
Assuming  that the spectra are 
the same (\ie~standard $\nu_e$), one gets for the 
$^8$B neutrino flux the results shown in 
Table 1. In the same table we also show the weighted average of
Kamiokande and Superkamiokande results.

\begin{table}%[t]
%\label{tabexp}
\vspace{0.2cm}
\begin{center}
\footnotesize
\caption[exper]{
 The main characteristic of each neutrino experiment:
 type, detection reaction, energy 
threshold $E_{th}$, experimental results with  statistical and
systematical errors. In the last column the  predictions of BP95 solar
model 
are presented \cite{BP95}. Errors are at $1\sigma$ level}
\begin{tabular}{|cccccc|}
\hline
Experiment& type & $E_{th}^{a)}$& result$^{b)}$ & combined$^{b)}$ &BP95$^{b)}$\\
\hline
&&&&&\\
Homestake & radiochemical & 0.814& 2.54 $\pm0.14\pm0.14$& 
          &9.3$^{+1.2}_{-1.4}$\\
          &$\nu+^{37}$Cl$\rightarrow$ e$^- + ^{37}$Ar&&&&\\
           \hline
            &&&&&\\
KAMIOKANDE & $\nu+ e^{-}\rightarrow \nu+ e^{-}$ & 7&$(2.80\pm0.19 \pm0.33)$& 
%$2.80 \pm 0.38$
          &\\
%          &&&&&\\ 
%          \hline
          &&&&&\\ 
          &&&&$2.51 \pm 0.16$& 6.62$(1.00^{+0.14}_{-0.17})$\\
          &&&&&\\
SUPERKAM.& $\nu+ e^{-}\rightarrow \nu+ e^{-}$& 6.5&$(2.44\pm0.06 ^{+0.25} _{-0.09})$& 
%$2.44 \pm 0.18$
          &\\
          &&&&&\\ 
          \hline
           &&&&&\\
GALLEX    & radiochemical &0.233&$76.2\pm6.5 \pm 5$&
% $76.2 \pm 8.2 $
           &\\
           &$\nu+^{71}$Ga$\rightarrow$ e$^- + ^{71}$Ge&&&&\\ 
%           \hline
            &&&&&\\
            &&&& $75 \pm 7$ & 137$^{+8}_{-7}$\\
	    &&&&&\\
SAGE      & radiochemical &0.233&$72^{+12 \,\, +5} _{-10 \,\, -7} $& 
% $72 \pm 13$
          &\\
           &$\nu+^{71}$Ga$\rightarrow$ e$^- + ^{71}$Ge&&&&\\
%           \hline
           \hline
\multicolumn{6}{|l|}{\scriptsize $^{a)}$ Energy in Mev}\\         
\multicolumn{6}{|l|}{\scriptsize $^{b)}$ in SNU for radiochemical exp.; 
in 10$^6$ cm$^{-2}$s$^{-1}$ for electron scattering exp.}\\          
\hline
\end{tabular}          
\end{center}
\end{table}
\vspace{0.2cm}

 All other experiments use 
radiochemical techniques. The $^{37}$Cl experiment of Davis and
coll.~\cite{Davis} has been the first operating solar neutrino detector. The
reaction used for neutrino detection is the one 
proposed by Pontecorvo in 1946\cite{Pontecorvo1946}:
\begin{equation}
	\nu_e+ {}^{37}{\mbox{Cl}} \rightarrow e^- +  {}^{37}{\mbox {Ar}} \, .
\end{equation}    

The energy threshold being 0.814 MeV, the experiment is 
sensitive mainly to \oB neutrinos, but also to \sBe neutrinos. The 
target, containing $10^5$ gallons of perchloroethylene, is located in the
Homestake gold mine in South Dakota.
Every few months a small sample of $^{37}$Ar (typically some fifteen atoms!)
was extracted from the tank and these radioactive atoms are counted in 
low background proportional counters. The result, averaged over more than 20
years of operation, is reported in Table 1.
The theoretical expectation is higher by a factor three.  For almost 20
years this discrepancy has been known as the ``Solar Neutrino Problem". 
 About 75\% of the total theoretical rate
is due to  \oB neutrinos and 
hence it was for a long time believed that the
discrepancy was due to the difficulty in predicting this rare source.

Two radiochemical solar neutrino experiments using $^{71}$Ga have given 
data:
GALLEX \cite{Gallex}, located at the Gran Sasso laboratory in Italy and using 30 tons 
of Gallium in an aqueous solution, and SAGE \cite{Sage}, in the Baksan valley 
in Russia, which uses 60 tons of gallium metal. The neutrino 
absorption reaction is
\begin{equation}
	\nu_e+^{71} {\mbox{Ga}} \rightarrow e^- + ^{71}{\mbox{Ge}} \, .
\end{equation}  

The energy threshold is $E_{th}=0.233$~MeV, 
and most of the signal arises from pp neutrinos with a significant
contribution  from \sBe neutrinos as well.
The Germanium atoms are removed chemically from the Gallium and the 
radioactive decays of  $^{71}$Ge  (half-life=11.4 days) are detected in small
proportional counters.  The results of the two experiments can be combined,
 see Table 1,
and we use the weighted average as 
representative value of the Gallium signal.
Again the value is almost a factor two below the theoretical prediction.

An overall efficiency test of the GALLEX  and SAGE
detector has been  performed by using  intense $^{51}$Cr neutrino 
sources \cite{Gallex,Sage}.
The number of observed neutrino events agrees with expectation to the
10\% level. This result
 ``provides an overall check of Gallium detectors, indicating that there are no
significant experimental artifacts or unknown errors at the 10\% level that 
are comparable to the 40\% deficit of observed solar 
neutrino signal''\cite{calibration}.

\section{Model independent analysis}
\label{secmodel}

The principal aim of this section is to extract information on the fluxes of 
solar  neutrinos directly from solar neutrino experiments, with 
minimal assumptions about solar models, see also Refs.
\cite{AA,PRD,WHERE,HATA1,HATA2,TAUP95,Report}

\subsection{Where are Be and CNO neutrinos?}

We make the assumption of stationary Sun (\ie ~the presently observed  
luminosity equals  the  present nuclear energy production rate) and 
standard neutrinos, so that  all the  $\nu_e$ produced in the Sun reach Earth 
without being lost and  their energy spectrum is unchanged.  The relevant 
variables are thus the (energy integrated) neutrino fluxes, 
which can be grouped as:
\begin{equation}
\fipppep, \quad \fibe, \quad \ficno \quad {\mbox{and}}\quad \fib \quad.
\end{equation}
These four variables, see~\cite{PRD,WHERE,Report}, are constrained by four 
relationships:\\

\noindent
a) the luminosity equation, implying that  the fusion of four protons  
(and two electrons) into one $\alpha$ particle is accompanied by the emission 
of two neutrinos, whichever is the cycle:
\begin{equation}
\label{lum}
   K_{\odot}=
\sum_i \left( \frac{Q}{2} - \langle E \rangle _i \right )  \Phi_i
\end{equation}
where $K_{\odot}$ is the solar constant 
($K_{\odot}=8.533\cdot 10^{11}$ MeV cm$^{-2}$ s$^{-1}$), Q=26.73 MeV and 
$\langle E \rangle _i $ is the average energy of the i-th neutrinos.\\

\noindent
b)The Gallium signal $S_G$=$(75\pm7)$ SNU 
can be expressed as a  linear combination of 
the $\Phi_i$'s, the weighting factors $\sigma_{i,G}$ being the absorption cross 
section  for the i-th neutrinos, averaged on their energy spectrum, 
see \eg~ Ref. \cite{Report} for updated values:
\begin{equation}
\label{sga}
S_G= \sum_i \sigma_{i,G} \Phi_i
\end{equation}

\noindent
c)A similar equation holds for the Chlorine 
experiment, $S_C=(2.54\pm0.20)$SNU:
\begin{equation}
\label{scl}
S_C= \sum_i \sigma_{i,C} \Phi_i
\end{equation}

\noindent
d)The KAMIOKANDE and SUPERKAMIOKANDE experiment determine
 - for standard neutrinos - the flux 
of Boron neutrinos:
\begin{equation}
\label{ska}
     \fib= (2.51\pm0.16)\cdot 10^6  {\mbox{cm$^{-2}$s$^{-1}$}}.
\end{equation}

\begin{figure}[t]
%\begin{center}
\epsfig{figure=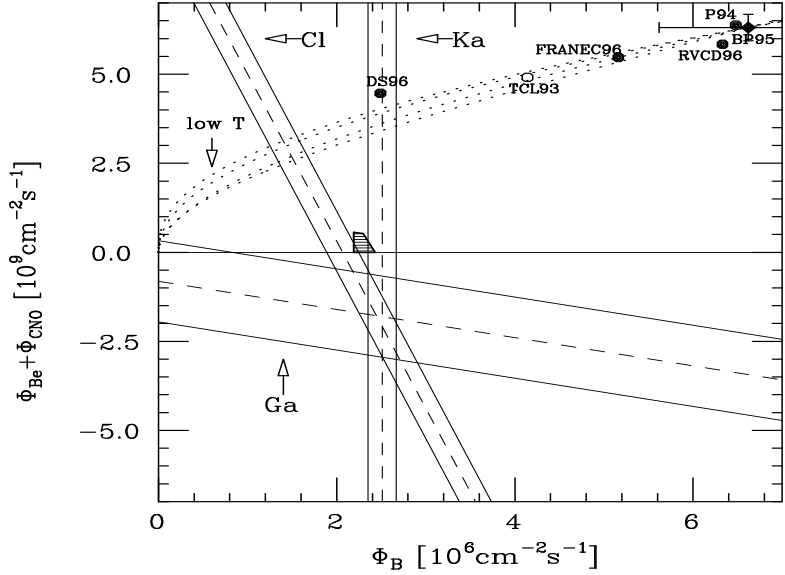,width=9cm,height=15cm,angle=90}
\caption[aff]{
The $^8$B and $^7$Be+CNO neutrino fluxes, 
consistent with the luminosity constraint
and experimental results for standard neutrinos.
The dashed (solid) lines correpond to the central ($\pm 1\sigma$) experimental
values for chlorine, gallium 
and $\nu - e$ scattering  experiments. 
The dashed area corresponds to the physical region within $2\sigma$ from each
experimental result.
The predictions of solar models including element diffusion (full circles)
~\cite{DS96,P94,BP95,RVCD96,Ciacio},
 and neglecting diffusion (open circles)  \cite{TCL93} 
are also shown.
The dotted lines indicate the behaviour of non standard solar models
with low central temperature.  }
\label{figbebo}
%\end{center}
\end{figure}

In order to understand what is going on, and to  make clear the role of 
each experimental result, let us reduce the number of equations and of 
unknowns by the following tricks:\\
(a) one can eliminate $\fipppep$ by using the luminosity 
    equation~(\ref{lum});\\
(b) since $\langle E\rangle _{CNO} \geq \langle E \rangle _{Be} $, 
 the corresponding cross section has to be larger 
than that of Be neutrinos. Thus the minimal CNO signal is obtained with 
the replacement
$\sigma_{CNO}\rightarrow \sigma_{Be}$.

In this way, the above equations can be written in terms of two variables, 
$\fibecno$ and $\fib$,  and the results of each experiment can be plotted in 
the ($\fib$, $\fibecno$) plane, see Fig.~\ref{figbebo}.

Clearly all  experiments point towards $\fibecno < 0$. This means that 
the 
statement ``{\em neutrinos are standard and experiments are correct}'' 
has lead us 
to an unphysical conclusion. Could the problem be with some experiment? It 
is clear from Fig.~\ref{figbebo} that the situation is unchanged by 
arbitrarily disregarding one of the experiments, see also Ref.~\cite{wrong}.

\subsection{Reduced central temperature models}

From Fig. \ref{figbebo} one easily understands why a 
reduction of the central solar
temperature $T$ cannot solve the solar neutrino puzzle.

\begin{figure}[t]
\epsfig{figure=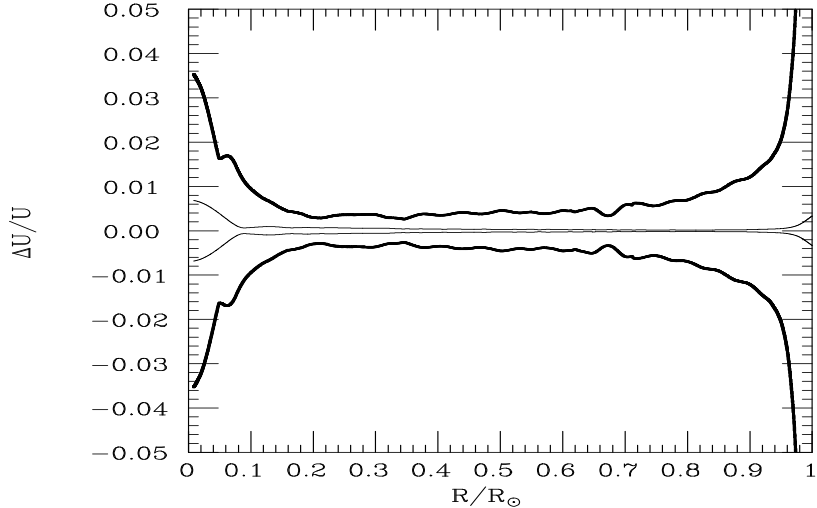,height=15cm, width=8cm, angle=90}
\caption[aaa]{
The estimated global relative uncertainty on $U=P/\rho$
(thick line) and that due to
the observational errors (thin line).
             }
 \label{fig1}
\end{figure}

Non standard solar models with smaller central temperaure can be obtained 
by varying -- well beyond the estimated uncertainties -- a few  parameters  
(the cross section of the pp reaction,  chemical composition, opacity, 
age... \cite{PRD,PLB}). These models span the dotted area in 
Fig.~\ref{fig1}, 
which can be clearly understood by simple considerations. 
As well knwon, all fluxes have approximately a power law dependence
on the central temperature \cite{Bahcall1989,PRD,PLB,Bahnew,Report}:
\begin{eqnarray}
\label{flux}
\fibe &=&\Phi_{\mbox{\footnotesize Be},0} (T/T_0)^{10}  \nonumber \\
\fib  &=&\Phi_{\mbox{\footnotesize B},0}   (T/T_0)^{20} \nonumber \\
\ficno&=&\Phi_{\mbox{\footnotesize CNO},0} (T/T_0)^{20}
\, ,
\end{eqnarray} 
where the subscript 0 refers here and in the following to the SSM 
predictions.
By expressing the temperature as a function of $\fib$, one has:
\begin{eqnarray}
\fibe + \ficno &=&
\Phi_{\mbox{\footnotesize Be},0} (\fib /
 \Phi_{\mbox{\footnotesize B},0})^{1/2} \nonumber \\
             &+&
\Phi_{\mbox{\footnotesize CNO},0} (\fib / \Phi_{\mbox{\footnotesize B},0}) 
\, ,
\end{eqnarray}
and one sees in Fig.~\ref{figbebo} the square root behaviour at small $\fib$,
 which changes to linear for larger $\fib$.
 
It is clear that all these model fail  to reproduce the experimental 
results, essentially because they cannot reproduce the observed ratio 
$\fibe / \fib $, see also Ref.~\cite{Bere,Report}.

\subsection{Conclusions}

In summary, we have  demonstrated that, under the assumption of standard
neutrinos:

\begin{itemize}
\item the available experimental results look inconsistent among themselves,
even if one of the experiments were wrong;

\item the flux of intermediate energy neutrinos (Be+CNO) as derived from
experiments is significantly smaller than the prediction of SSM's;

\item the different reduction factors for $^7$Be and $^8$B neutrinos
with  respect to the SSM are essentially in
contradiction with the fact that both $^7$Be and $^8$B neutrinos
originate from the same parent $^7$Be nucleus.
\end{itemize}

\section{Implications of helioseismology}
\label{sechelios}

Helioseismology allows us to look into the deep interior of the Sun,
probably more efficiently than with neutrinos. The highly precise 
measurements of frequencies and the tremendous number of 
measured lines  enable us to extract the values of sound 
speed inside the sun with accuracy better than 1\%.  Recently it 
was demonstrated that a comparable accuracy can be obtained for 
the inner solar core \cite{eliosnoi}.

In this section we summarize the results of our group in the last
year concerning a systematic analysis of helioseismic implications
on solar structure and neutrino production.
We quantitatively estimated  the  accuracy
of solar structure properties
as inferred from the measured frequencies
through the so called inversion method. This analysis provided 
the base for quantitative tests of solar models. These
tests are briefly presented  here, see  \cite{eliosnoi,eliospp,eliosmix}
for more details.

\subsection{How accurate are solar properties as inferred from
helioseismology?}

We remind that  solar observations provide measurements of
the frequencies  
\{$\nnu$\} of solar p-modes, and  quantities $Q$ characterizing the solar 
structure are indirectly inferred from the \{$\nnu$\}'s, through an inversion 
method.  Schematically, the procedure is the following:

a)One starts with a solar model, giving values $Q_{mod}$ 
 and predicting a set  \{$\nnu_{mod}$\} of frequencies.  
These  will be somehow different from the measured frequencies, 
$\nnu_\odot \pm \Delta \nnu_\odot$.

b)One  then searches for  the corrections  $\wDelta Q$ to the solar model which 
are needed in order to match the corresponding frequencies
 \{$\nnu_{mod} + \oomega$\}  with the observed 
frequencies \{$\nnu_{\odot}$\}. 
Expression for $\oomega$ are derived 
 by using 
perturbation theory, where  the starting model is used as a zero-th order 
approximation. 
The correction factors $\wDelta Q$ are then computed, assuming some regularity
 properties, so that
 the problem is  mathematically well defined and/or 
unphysical solutions are avoided.

c)The ``helioseismic value'' $Q_\odot$ is thus determined by adding 
the starting value and the correction 
\footnote{
Concerning notation, we remark that $\wDelta Q$ indicates the correction to
the starting solar model so as to obtain helioseimic value (see Eq. \ref{corr}),
 whereas $\Delta Q$ indicates
the estimated uncertainty on $Q$.}:
\begin{equation}
\label{corr}
Q_\odot= Q_{mod} + \wDelta Q  \, .
\end{equation}

For each quantity $Q$  we have determined the partial errors 
corresponding 
to each  uncertainty of the helioseismic method.
In fact, there are three independent sources of errors in the 
inversion process:

i)Errors on the measured frequencies, which -- for a given inversion 
procedure -- propagate on  the value of $Q_\odot$.

ii)Residual dependence on the starting model:
the resulting $Q_\odot$ is slightly different
if one starts with  different   solar models. This introduces an 
additional uncertainty, which can be evaluated by 
comparing the results of several calculations.

iii)Uncertainty in the regularization procedure. Essentially this is a 
problem of 
extrapolation/parametrization. Different methods, equally acceptable in 
principle, yield (slightly) different values of $Q_\odot$.

It has to be remarked that, in view of the extreme precision of the measured 
frequencies\cite{ref1,ref2,ref3,ref4},
 $ \Delta \nnu_{\odot} / \nnu_{\odot} \lapprox 10^{-4}$, 
uncertainties corresponding to ii) and iii) are extremely important.

For deriving a global uncertainty, we took a very
conservative approach.  May be that the parameter variation was not 
exhaustive, and what we found as extrema are not really so, but actually are 
quite acceptable values. In view of this, let us double  the interval we found 
and interpret $\pm( \Delta Q  )_k$, as partial errors. Furthermore, let us be 
really {\em conservative} 
assuming that errors add up linearly. In conclusion, this gives:
\begin{equation}
\label{cons}
\Delta Q =  \pm  \sum_k  | (\Delta Q)_k |  \, .
\end{equation}

 With this spirit, we  analysed  
several physical quantities $Q$ characterizing the solar structure.
Concerning the outer part of the sun, we  discussed  the photospheric
helium abundance
Y$_{ph}$,  the depth of the the convective envelope $R_b$, 
and the density  at  its  
bottom   $\rho_b$.
Then  we  considered  the ``intermediate" solar interior ($x$=$R/R_\odot =0.2-
0.65$), analysing the behaviour of the squared isothermal sound speed,
\ug=$P/\rho$.
Finally we  investigated the inner region ($x \leq 0.2$), where 
nuclear energy  and neutrinos are produced.

\begin{table}%[ht]
\label{tabhelios}
\vspace{0.2cm}
\begin{center}
\caption[errori]{
For the indicated quantities $Q$ we present the helioseismic
values
$Q_\odot$ and the relative  errors from Eqs. (\ref{cons},\ref{stat}).
 All uncertainties are in  \permille. 
In the  fifth and sixth row, 
for  $U=P/\rho$ the values of 
 the uncertainties are the maxima in the 
indicated interval.
               }
\vspace{0.2cm}
\begin{tabular}{|lcccccccc|}
\hline
$Q$ &&&$Q_\odot$&&&$\left (\frac{\Delta Q}{Q}\right)$&& $\left (\frac{\Delta Q}{Q}\right)_{1\sigma}$ \\
\hline
Y$_{ph}$ &&&0.249&& &  42&& 14\\
$R_b/R_\odot$& &&0.711&&& 4 && 2 \\
$\rho_b$ [g/cm$^3$] &&& 0.192 &&& 37  && 9.4\\
%\hline
$U$($0.2<x<0.65$ )&&& &&& 5&& 1.4 \\
$U$($0.1<x<0.2$ )&&&  &&& 9.4&& 2.3  \\
%\hline
$U$(0) [$10^{15}$ cm$^2$ s$^{-2}$]&&& 1.54  &&& 35 &&  10  \\
%$U$$(x_{Be})$ [$10^{15}$ cm$^2$ s$^{-2}$]&&& 1.56  &&& 17  &&\\
%$U$$(x_{B})$  [$10^{15}$ cm$^2$ s$^{-2}$] &&& 1.56 &&&  22 &&\\
\hline
\end{tabular}
\end{center}
\end{table}

From Table 2 one can see that
the three independent physical 
properties of the convective envelope (Y$_{ph}$, $R_b$ and
$\rho_b$)
 are determined very accurately by seismic observations
\footnote{A fourth seismic ``observable", the sound speed at the convective radius
is traditionally considered, e.g. \cite{c-d}. We have not included it in 
our list 
since, as shown in Ref.\cite{elios}, it is not an independent one.}.
We remark that for all these quantities 
the uncertainty resulting from propagation 
of the frequency measurement errors
is of minor importance with respect to the 
``systematic'' errors, intrinsic to the 
inversion method \cite{eliosnoi}.

In the intermediate solar region, 
 the helioseismic
determination is extremely accurate:
$|\Delta U/U | \leq  5{\mbox{\permille}}$
throughout the explored region,
where most of the error again is from uncertainties in the inversion
method, see Fig. 2.

As well known, most of the energy and of solar neutrinos originate 
from the innermost part  of the sun. According to SSM calculations,
see \eg~Refs. \cite{Report,BP95}, about  
94\% of the solar luminosity  and 93\% of the pp neutrinos are produced 
within $x<0.2$.
Clearly the helioseismic precision worsens in this region, due to the fact
 that the observed p-modes do not penetrate 
in the solar core, and consequently the information one can
 extract from available experimental results is limited,
but still important.
Even at the solar centre, the accuracy is still 3.5\%.

 In conclusion,
{\em helioseismology provides significant insight even on the solar
innermost core}.
\begin{figure}[t]
\begin{center}
\epsfig{figure=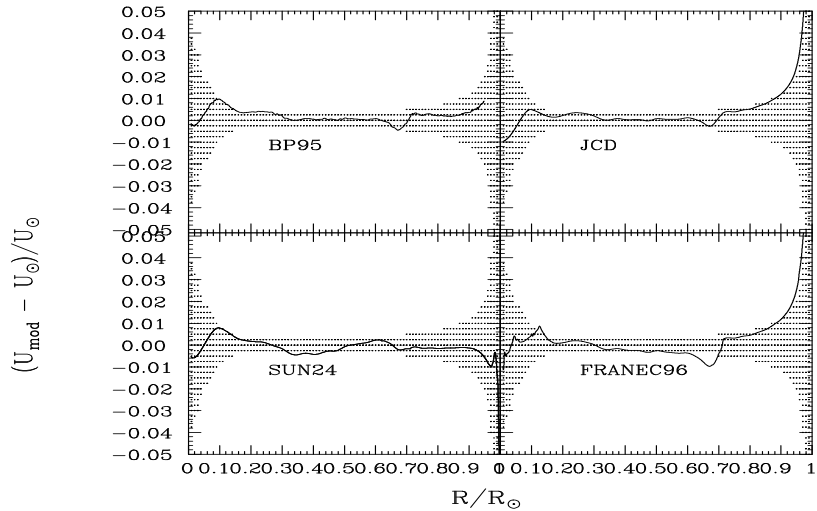,height=15cm, width=8cm, angle=90}
\caption[abb]{
The difference between $U$ as predicted by selected solar 
models, $U_{mod}$, and the helioseismic determination, $U_{\odot}$,
 normalized to this latter. 
The dotted area corresponding to 
$\left ( \frac{\Delta U}{U}  \right ) $.
SUN24 is the ``model 0'' of Ref. 
\cite{ref10};
FRANEC96 is the ``best'' model 
with  He and heavier elements diffusion of
Ref. \cite{Ciacio};  BP95 is the model with metal and He diffusion of 
Ref. \cite{BP95}; JCD is the 
``model S'' of Ref. \cite{JCD}.             }
 \label{fig2}
\end{center}
\end{figure}

Concerning error estimate,
we remark that we have been extremely conservative using
Eq. (\ref{cons}), which should provide a sort of reliable
``$3\sigma$ error''. By combining the partial errors in quadrature,
the resulting global error
\begin{equation}
\label{stat}
(\Delta Q)_{1\sigma} =  \pm \frac{1}{2} \sqrt { \sum_k  (\Delta Q)_k ^2 }
\end{equation}
is clearly reduced, see again Table 2. This estimate is similar to
that quoted in Ref. \cite{BPBCD}.

\subsection{Helioseismology and SSMs}

\begin{figure}[t]
\begin{center}
\epsfig{figure=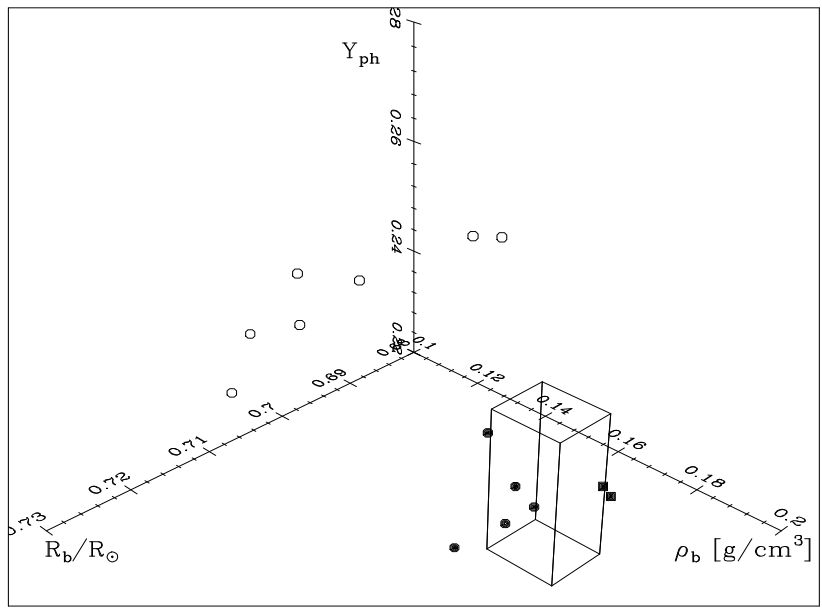,height=13cm, width=9cm, angle=90}
\caption[aaa]{
Helioseismic determinations and solar model predictions about
the convective envelope.
The box defines the region allowed by helioseismology.
Open circles denote models without diffusion, squares models with He diffusion,
full circles models with He and heavier elements diffusion, see Ref.
\cite{eliosnoi}.
             }
\label{fig3}
\end{center}
\end{figure}

The comparison between the predictions of a few recent SSM
calculations and helioseismic information is shown in Figs. \ref{fig2}
and \ref{fig3}.

Concerning the (isothermal) sound speed profile, see Fig. \ref{fig2},
all models look generally
good. Also SUN24, a model which neglects 
elemental diffusion, passes this test.
The study of the  convective envelope is illuminating, see Fig. \ref{fig3}.
 {\em All models neglecting  
diffusion are in clear contradiction with helioseismic constraint. On the other
hand, calculations where diffusion is included look in  substantial agreement
with helioseismology.}
All this shows that the two approaches  (profile of $U$ and properties of the 
convective envelope) are complementary and both important.

The previous arguments show that SSMs are in good shape. 
Actually, helioseismology provides a  new perspective/definition of SSMs.
Before the advent of helioseismology a SSM had 
{\em three} essentially free parameters,
$\alpha$, Y$_{in}$ and (Z/X)$_{in}$ for producing {\em three} measured
quantities: the present radius, luminosity and heavy element 
content of the photosphere. This may not look as a too big accomplishment,
in itself.
Nowadays, {\em by using the same number of parameters a SSM has to 
reproduce many additional data}, such as 
Y$_{ph}$, $R_b$, $\rho_b$, $U(R)$, provided
by helioseismology.

Alternative solar models have to be 
confronted with these data too, see \eg~ Refs. \cite{eliospp,eliosmix}.

\begin{table}
\caption[abc]{
Predictions for neutrino fluxes and signals in the Cl and Ga detectors 
from HCSMs. Uncertainties corresponding
 to $(\Delta T/T )=\pm 1.4\%$ 
are shown  together with those from
 nuclear cross sections at 3$\sigma$.
The $1\sigma$ estimated error (see text) is shown in the last
column.
}
\vspace{0.2cm}
\begin{center}
\begin{tabular} {|ll |c |c| c | c  |}
\hline
  && $Q_{HCSM}$ & $\Delta Q _T$ & $\Delta Q _{nuc}$  & $\Delta Q _{1\sigma}$\\
\hline
$\fibe$ &$[10^9$/cm$^{2}$/s$]$ &  4.81&$\pm 0.53$& $\pm 0.59$ & $\pm$0.3\\
 $\fib$ &$[10^6$/cm$^{2}$/s$]$ &  4.90&$\pm 1.22$ &$\pm 0.94$ & $\pm$0.5 \\
 Cl &$[$SNU$]$& 7.2&$\pm 1.7$&$\pm 1.2$   & $\pm$0.7\\
Ga &$[$SNU$]$&  130&$\pm 10$ &$ \pm 7$ & $\pm$4\\
\hline
\end{tabular}
\end{center}
\label{tabhcsm}
\end{table}

\subsection{Helioseismically constrained solar models and neutrino
fluxes}

 Actually, one can exploit helioseismology within a 
 different strategy. One can relax some assumptions 
 on the most controversial ingredients of solar models
  (\eg~ opacity and metal abundance) and determine them 
  by requiring that helioseismic constraints are satisifed. 
  These Helioseismically Constrained Solar Models (HCSM)
   all yield  the same central temperature within  about 
   one percent $T=1.58 \times 10^7$K, see Ref. \cite{HCSM}.
Depending on the error definition (see Eqs. (\ref{cons},\ref{stat}))
one has:
\begin{equation}
\label{tempera}
\Delta T/ T = \pm 1.4\% \quad \quad  (\Delta T/T)_{1\sigma}= \pm 0.5\%
\end{equation}

The predicted neutrino fluxes and signals are shown in 
Table \ref{tabhcsm}, which is obtained by using the most
recent determination of nuclear cross sections, in particular we use now
$S_{17}=(18.4\pm0.9)$eV b \cite{s17}.  This table updates
the  corresponding one in Ref. \cite{HCSM}.

Note that the astrophysical uncertainties (calculated
with the conservatively estimated $\Delta T/T$) and 
the nuclear physics uncertainties (estimated as $3\sigma$
 errors) are now quite similar.
The same conclusion holds if one uses $(\Delta T/T )_{1\sigma}$
together with $1\sigma$ errors on nuclear cross sections.
Of course, there is some freedom about how to combine
astrophysical and nuclear physics uncertainties. As
an example, the result of adding in quadrature the effect
of $(\Delta T )_{1\sigma}$ and of $1\sigma$ nuclear
cross section errors is shown in the last column of Table 
\ref{tabhcsm}.

\subsection{Conclusions}

As a summary of this section, let us outline the main points:

\begin{itemize}

\item
Helioseismology provides significant information about the solar
structure, even at the solar innermost core, a conservative error
estimate giving: $\Delta U/ U (0)=3.5\%$.

\item
Recent SSMs calculations, including element diffusion, are in agreement with
helioseismology.

\item
Helioseismically constrained solar models predict the central solar temperature
with a ``$1\sigma$'' accuracy of 0.5\%. This result is essentially
independent on uncertainties of solar opacity, which is used as a free
parameter, fixed by helioseismic results.

\end{itemize}

\begin{table}
\caption{The proposed solutions, their fingerprints and the experiments
looking at them.}
\begin{center}
\footnotesize
\begin{tabular}{|lccccc|}
\hline
Proposed solution & \multicolumn{5}{c|}{Signatures}\\
\hline
 & Oscillation & Spectral & Day-night &  Seasonal & CC/NC \\
 & at reactor  & deformation & variation & modulation & events \\
 \hline
 MSW small angle & NO & TINY & TINY & NO & YES \\
 MSW large angle & NO & TINY & YES & NO & YES \\
 JUST-SO         & NO & YES & NO & YES & YES \\
Universal oscil. & YES & NO & NO & NO & YES \\
\hline
Experiment        & CHOOZ & SUPERKAM. & SUPERKAM. & BOREXINO & SNO \\
Data             & now & now & now & 2000 & 2000 \\
\hline
\end{tabular}
\end{center}
\label{Fiorentini.tab1}
\end{table}

\section{Future prospects}
\label{secfuture}

\begin{figure}[t]
\begin{center}
\epsfig{figure=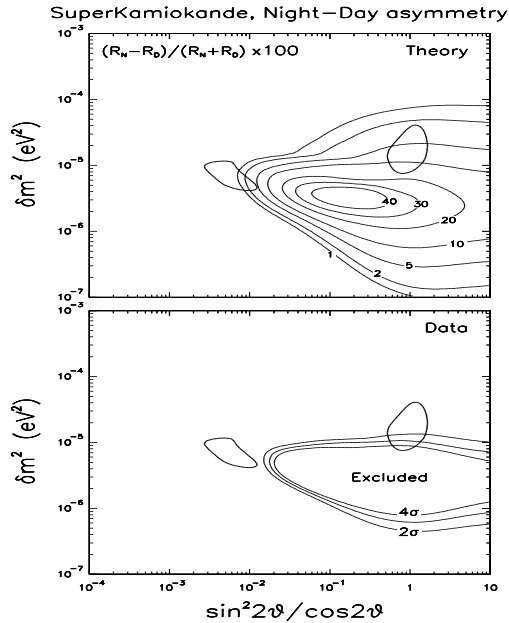,height=9cm, width=10cm}
\caption[aaa]{
Constraint on the MSW solutions as derived by the measured
value of the asymmetry of nihttime ($R_N$) and daytime ($R_D$) 
event rates. Upper panel: theoretical expectations for 
$(R_N-R_D)/(R_N+R_D)$. Lower panel: Regions excluded at 2, 3 and
4 standard deviations by the SUPERKAMIOKANDE data.
(From Ref. \cite{lisi}.)
             }
\end{center}
\label{figsk6}
\end{figure}

In the prophetical paper of 1946 \cite{Pontecorvo1946}
 Bruno Pontecorvo wrote:
``direct proof of the  {\em existence} of the neutrino ... 
must be based on experiments the interpretation
 of which does not require the law of conservation of 
 energy, i.e. on experiments in which some characteristic 
 process produced by free neutrinos ... is observed''.

\begin{figure}[ht]
\begin{center}
\epsfig{figure=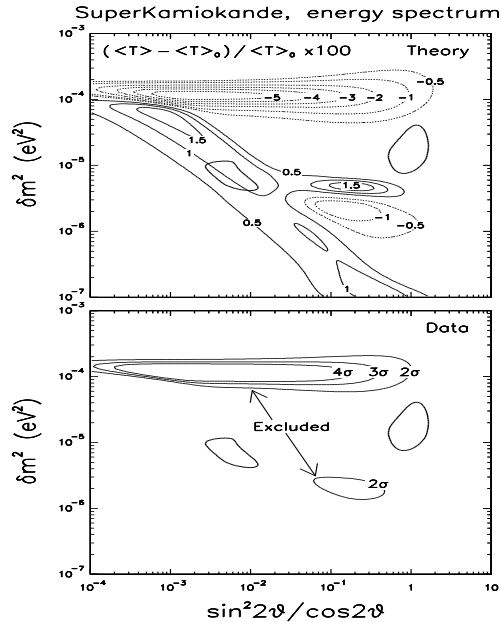,height=9cm, width=10cm}
\caption[aaa]{
Constraints on the MSW solutions as derived by the measured value
of the average electron kinetic energy $<T>$.
Upper panel: theoretical expectations for the fractional shifts
of $<T>$from its standard (no oscillation)
value $ <T>_o$.Lower panel: region excluded
at 2, 3, and 4 standard deviations by SUPERKAMIOKANDE determination
of $<T>$.  
(From Ref. \cite{lisi}.)
             }
\end{center}
\label{figsk5}
\end{figure}

The situation now looks very similar, just change 
{\em existence} with  {\em non standard  properties},
 in that the strongest argument for a particle physics
 solution to the SNP arises from energy conservation 
 (the luminosity constraint) and actually we need a direct 
 footprint of some neutrino property, not predicted within the minimal
  standard model of electroweak interactions.

\begin{figure}[ht]
\begin{center}
\epsfig{figure=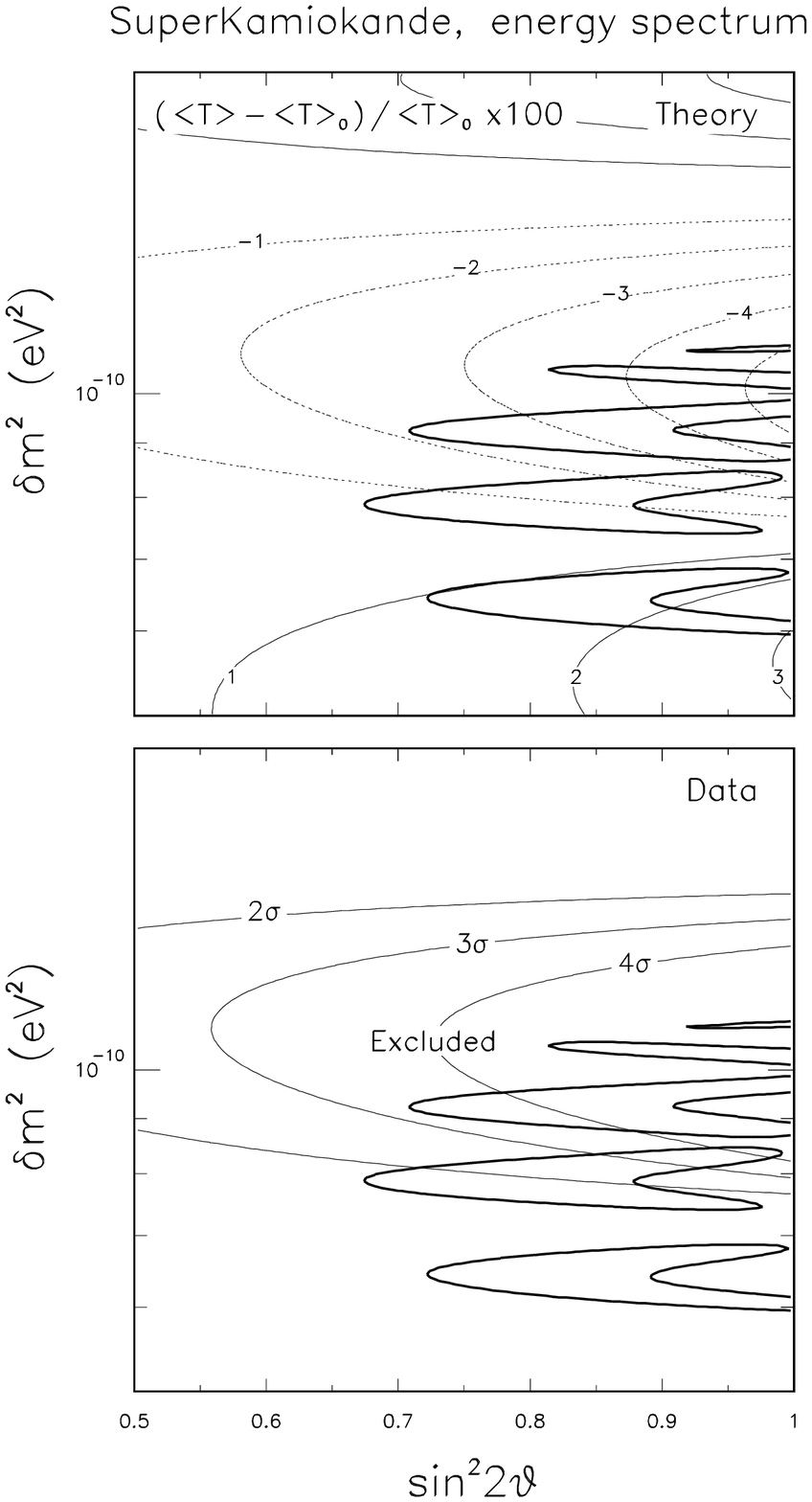,height=9cm, width=10cm}
\caption[aaa]{
Constraints on the vacuum oscillation
 solutions as derived by the measured value
of the average electron kinetic energy $<T>$.
Upper panel: theoretical expectations for the fractional shifts
of $<T>$ from its standard (no oscillation)
value $<T>_o$. Lower panel: region excluded
at 2, 3, and 4 standard deviations by SUPERKAMIOKANDE determination
of $<T>$.  
(From Ref. \cite{lisi}.)
             }
\label{figsk7}
\end{center}
\end{figure}

The four  most popular particle physics solutions 
(small and large angle MSW effect, just so oscillations 
and universal oscillations) all predict specific signatures
 which are being or will be tested by the new generation 
 of experiments (Superkamiokande, Borexino, SNO...), 
 see Table \ref{Fiorentini.tab1}.
 
The universal oscillation solution particularly substained 
by Perkins \etal~ \cite{Fiorentini.10} ($\Delta m^2 \approx 10^{-2}$ eV$^2$
and threefold maximal mixing) might account for  both the KAMIOKANDE
atmospheric neutrino anomaly and the results of solar neutrino experiments
(possibly with some stretching of error bars).

This hypothesis has just been falsified by the recent negative 
 result of CHOOZ \cite{Fiorentini.11}. This nice and small 
 (in comparison with the gigantic solar neutrino devices) 
 experiment at a nuclear reactor is cleaning  some of the fog in the air.

The large angle MSW solution ($\Delta m^2 \approx 10^{-5}$ eV$^2$, twofold 
maximal mixing) is being under the attack of SUPERKAMIOKANDE
\cite{lisi}.
The absence so far of a day-night modulation
\begin{equation}
\frac{N-D}{N+D} = 0.017\pm 0.026 \, {\mbox{(stat.)}} \, 
\pm 0.017  \, {\mbox{(syst.)}} 
\end{equation}
clearly excludes a significant portion of the parameter
space for this solution, see Fig. 5 and Ref. \cite{lisi}.
On the other hand, the electron spectrum deformations
are very small \cite{vignaud,lisi}, see Fig. 6, and hardly
detectable with SUPERKAMIOKANDE.

The small angle MSW solution 
($\Delta m^2 \approx 10^{-5} $ eV$^2$, $sen^2 2\theta \approx 10^{-(2\div3)}$)
is perhaps the most elusive one. As shown 
in Figs. 5 and 6 both day-night effect and spectrum
deformations are very tiny and can escape to SUPERKAMIOKANDE.
Possibly, the only detectable signature in future esperiments is in
the ratio of Charged Current to Nuclear Current events
(CC/NC), see below.

The Just So solution ($\Delta m^2 \approx 10^{-10}$ eV$^2$ and maximal mixing)
implies spectral deformation which can be detected with 
SUPEKAMIOKANDE (see Fig. \ref{figsk7}) and 
seasonal modulations of the \sBe signal which
will be the realm of BOREXINO \cite{borex,justso}.

The SNO experiment \cite{sno} looks to us, in
many respect, as the final remedy/last hope.
SNO will be capable of detecting neutral current events, which
are produced by any {\em active} neutrino.
The measurement of the active neutrino flux ($\Phi _e +\Phi_\mu$)
in the \oB energy region, combined with the SUPERKAMIOKANDE
information ($\Phi _e +1/7 \, \Phi_\mu$), and/or with the charghed current
signal of SNO ($\Phi _e$), can provide a define proof of neutrino oscillation.

This holds for  {\em any} of the proposed solutions, unless nature decides
that $\nu_e$ convert into sterile neutrinos.

Let us wait and wish that (at least) one of 
the fingerprints of neutrino oscillations  
is detected by the new experiments.

\section*{Acknowledgments}
We are extremely grateful to E. Lisi for useful comments and discussions.

\end{document}